\documentclass[epj]{webofc}
\usepackage[utf8]{inputenc}
\usepackage[varg]{txfonts}   
\usepackage{booktabs}
\usepackage{xcolor}
\definecolor{darkred}{rgb}{0.4,0.0,0.0}
\definecolor{darkgreen}{rgb}{0.0,0.4,0.0}
\definecolor{darkblue}{rgb}{0.0,0.0,0.4}
\usepackage[bookmarks,linktocpage,colorlinks,
    linkcolor = darkred,
    urlcolor  = darkblue,
    citecolor = darkgreen]{hyperref}
\usepackage{subfigure}
\wocname{EPJ Web of Conferences}
\woctitle{Lattice2017}
\def\be#1{\begin{equation}#1\end{equation}} 
\begin{document}
%
\selectlanguage{english}
\title{%
Dirac spectral density and mass anomalous dimension in 2+1 flavor QCD
}
\author{%
\firstname{Katsumasa} \lastname{Nakayama}\inst{1,2}\fnsep\thanks{\email{katumasa@post.kek.jp}} \and
\firstname{Shoji} \lastname{Hashimoto}\inst{2,3} \and
\firstname{Hidenori}  \lastname{Fukaya}\inst{4}\fnsep
}
\institute{%
Department of Physics, Nagoya University, Nagoya, 464-8602, Japan
\and
KEK Theory Center, 
  High Energy Accelerator Research Organization (KEK), 
  Tsukuba 305-0801, Japan
\and
School of High Energy Accelerator Science, 
  The Graduate University for Advanced Studies (Sokendai),
  Tsukuba 305-0801, Japan
 \and
 Department of Physics, Osaka University, Toyonaka, Osaka 560-0043 Japan
}
\abstract{%
We compute the Dirac spectral density of QCD in a wide range of eigenvalues by using a stochastic method. We use 2+1 flavor lattice ensembles generated with Mobius domain-wall fermion at three lattice spacings ($a=0.083, 0.055, 0.044$ fm) to estimate the continuum limit. The discretization effect can be minimized by a generalization of the valence domain-wall fermion. The spectral density at relatively high eigenvalues can be matched with perturbation theory. We compare the lattice results with the perturbative expansion available to $O(\alpha_s^4)$.
}
\maketitle
\section{Introduction}\label{intro}

Eigenvalue spectrum of the Dirac operator carries rich information of QCD dynamics.
A well-known example is the Banks-Casher relation \cite{Banks:1979yr}, which relates the chiral condensate, a consequence of non-perturbative QCD dynamics, to the low-lying eigenvalue density.
Through this relation, the chiral condensate has been calculated on the lattice and precise determination is achieved \cite{Cossu:2016eqs}.

In this work, we calculate the Dirac spectrum in the high energy region where the eigenvalue is much larger than the QCD scale $\Lambda_\mathrm{QCD}$.
The spectral function is mostly perturbative in this region and the perturbative coefficients are known up to $O(\alpha_s ^4)$.
In the past, the Dirac spectral density has been used to extract the mass anomalous dimension of many-flavor QCD, which is expected to exhibit conformal scaling \cite{Cichy:2013eoa,Patella:2012da,Cheng:2013eu}.
In this work we utilise the same quantity but for the case of non-vanishing $\beta$ function, in order to study the convergence of perturbation theory and to extract the strong coupling constant $\alpha_s$.

By generalizing of the Banks-Casher relation, one can write the Dirac spectral density in terms of the  chiral condensate at an imaginary mass \cite{Damgaard:1998xy},
\begin{equation}
\rho(\lambda) = - \frac{1}{2\pi}
\left(
\langle \overline{q}q(m=i\lambda+\epsilon)\rangle
-
\langle \overline{q}q(m=i\lambda-\epsilon)\rangle
\right),
\label{chiralcond}
\end{equation}
where the valence quark mass is set to $m = i\lambda$ when evaluating the expectation value of the scalar density operator $\overline{q}q$.
In perturbation theory, it is calculated to $O(\alpha_s ^3)$ \cite{Kneur:2015dda} as summarized in (\ref{eq:ordert}) below.

To calculate the Dirac eigenvalue denisity on the lattice, we use a stochastic method with a filtering function approximated by the Chebyshev polynomial. The number of eigenvalues $n[v,w]$ of hermitian matrix $A$ in a range $[v,w]$ can be estimated with a filtering function $h(x)$, which is 1 in the range $[v,w]$ and zero otherwise, as
\begin{equation}
  n[v,w]\simeq \frac{1}{N}\sum_{k=1}^N \langle\xi_k h(A)\xi_k\rangle,
\end{equation}
where $\xi_i$ is a normalized gaussian noise vector and $N$ is its number.
We approximate the filtering function $h(x)$ by a polynomial of the form $h(x) = \sum k_iT_i(x)$ where $T_i(x)$ is the Chebyshev polynomial and  $k_i$ is a coefficient uniquely determined depending on $[v,w]$. The Chebyshev polynomial may be constructed by a recursion relation $T_0(x) = 1$, $T_1(x) = 2x$, and $T_{i+1}(x) = 2xT_i(x) - T_{i-1}(x)$.
Once we calculate the inner products $\langle\xi_kT_i(x)\xi_k\rangle$, the bin size $[v,w]$ can be chosen afterwards, so that the whole spectrum is obtained by one pass.
The details are in \cite{Cossu:2016eqs,Napoli}.
Then, the Dirac spectral density can be written as
\be{
  a^3\rho(\lambda)
  =
  \frac{1}{2V/a^4}\frac{n[v,w]}{a\delta},
}
with lattice volume $V$, bin size $\delta$, and lattice spacing $a$. Using this relation, we can calculate the Dirac spectral density by identifying $A = 2a^2D^\dagger D - 1$, which is constructed by the Dirac operator $D$.

We need to carefully address the discretization effects since the high modes are sensitive to them, and they strongly depend on the fermion formulation.
We investigate the discretization effect for the domain-wall fermion formulation and  find that the effect on the eigenvalue largely depends on the parameters in the formulation.
In particular, by adjusting the value of the Pauli-Villars mass, the discretization error appearing in the Dirac eigenvalue is greatly reduced without sacrifysing the locality and chirality.
This generalization of the domain-wall fermion formulation is discussed in Section \ref{sec-1}.

Perturbative expansion of $\rho(\lambda)$ is another important element of this study.
We write down the expansion up to $O(\alpha_s ^4)$ using the renormalization group equation for the spectral density as described in Section \ref{sec-2}.
Finally we compare our lattice calculation with the perturbative calculation in Section \ref{sec-3}.

\section{Domain-wall fermion with generalised Pauli-Villars mass}\label{sec-1}

In the lattice calculation of the Dirac spectral density we try to reduce the artifact due to  discretized space-time.
Discretization effects become more significant when we consider the physics in the perturbative scale,
where the bulk of the effect originates from the fermion formulation. We use the Mobius domain-wall fermion in this work.

The domain-wall fermion may be considered as an implementation of the Ginsparg-Wilson relation with a particular approximation of the sign function.
It is defined in five dimensional Euclidean space, and the four-dimensional theory is obtained from its surface modes.
The details of the Mobius domain-wall fermion may be found in \cite{Brower:2012vk,Boyle:2015vda}.

The four-dimensional effective operator $D_{\mathrm{ov}}$ for the Mobius domain-wall fermion may be written as
\begin{equation}
  \label{eq:Dov}
  aD_{\mathrm{ov}}(m_f,m_p) =
  (2-(b-c)M_0)M_0m_p
  \frac{(1+m_f)+(1-m_f)\gamma_5 \mathrm{sgn}(\gamma_5 aD_M)}{
        (1+m_p)+(1-m_p)\gamma_5 \mathrm{sgn}(\gamma_5 aD_M)},
\end{equation}
with a fermion mass $m_f$, the Pauli-Villars mass $m_p$, and ``$\mathrm{sgn}$'' an approximated sign function. Here, the Mobius kernel operator $D_M$ is defined as
\begin{equation}
  \label{eq:kernel}
  aD_M = \frac{(b+c) aD_W}{2 + (b-c) aD_W},
\end{equation}
in terms of the Wilson-Dirac operator $D_W$, which is reduced to
\begin{equation}
aD_\mathrm{W} = i\sum_{\mu}{}\gamma_\mu\mathrm{sin}ap_\mu + r\sum_{\mu}{} (1 - \mathrm{cos}ap_\mu) - M_0,
\end{equation}
in the momentum space when the background gauge field is absent.

\begin{figure}[thb]
  \centering
  \sidecaption
  \includegraphics[width=4.5cm, angle=-90]{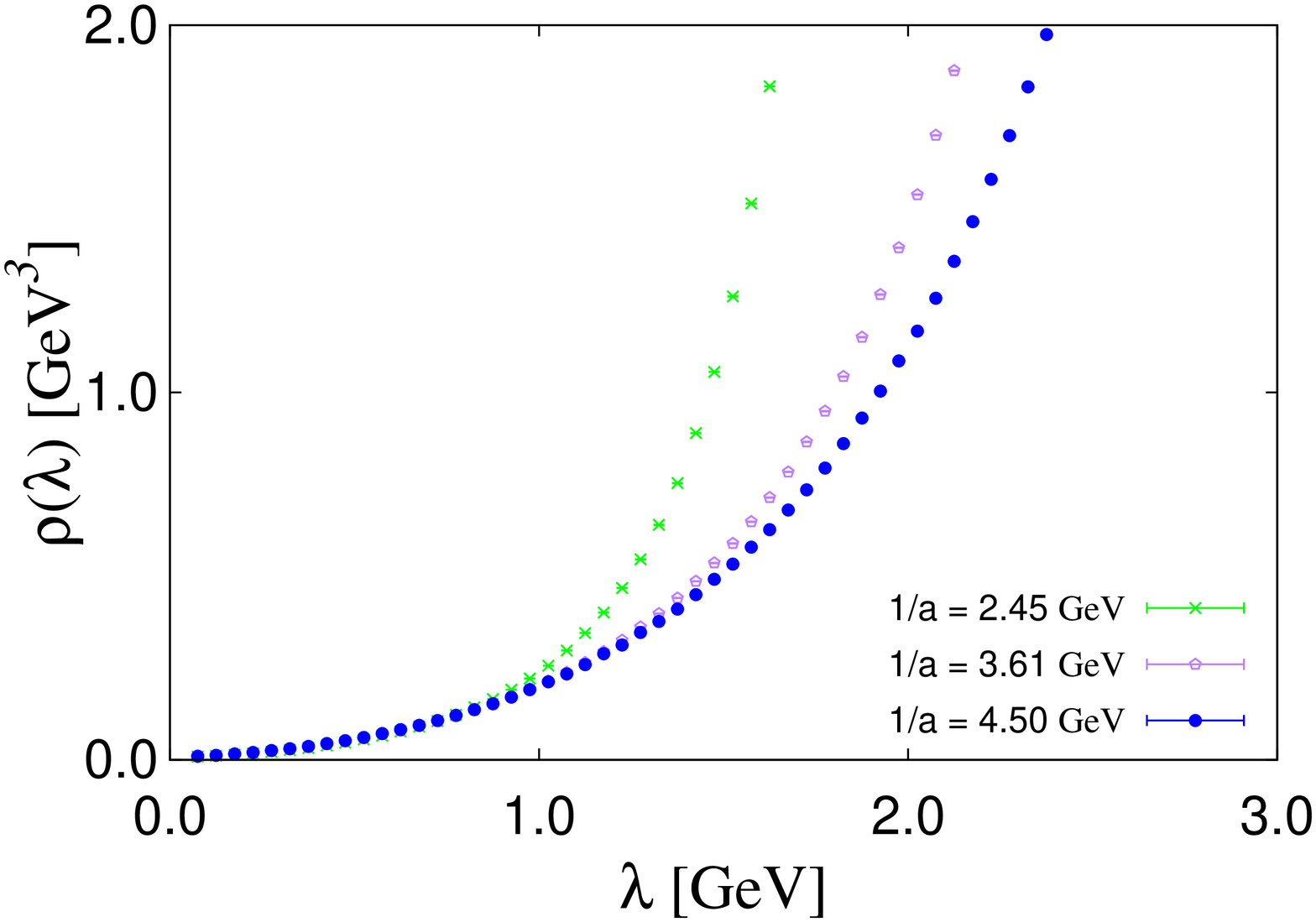}
  \includegraphics[width=4.5cm, angle=-90]{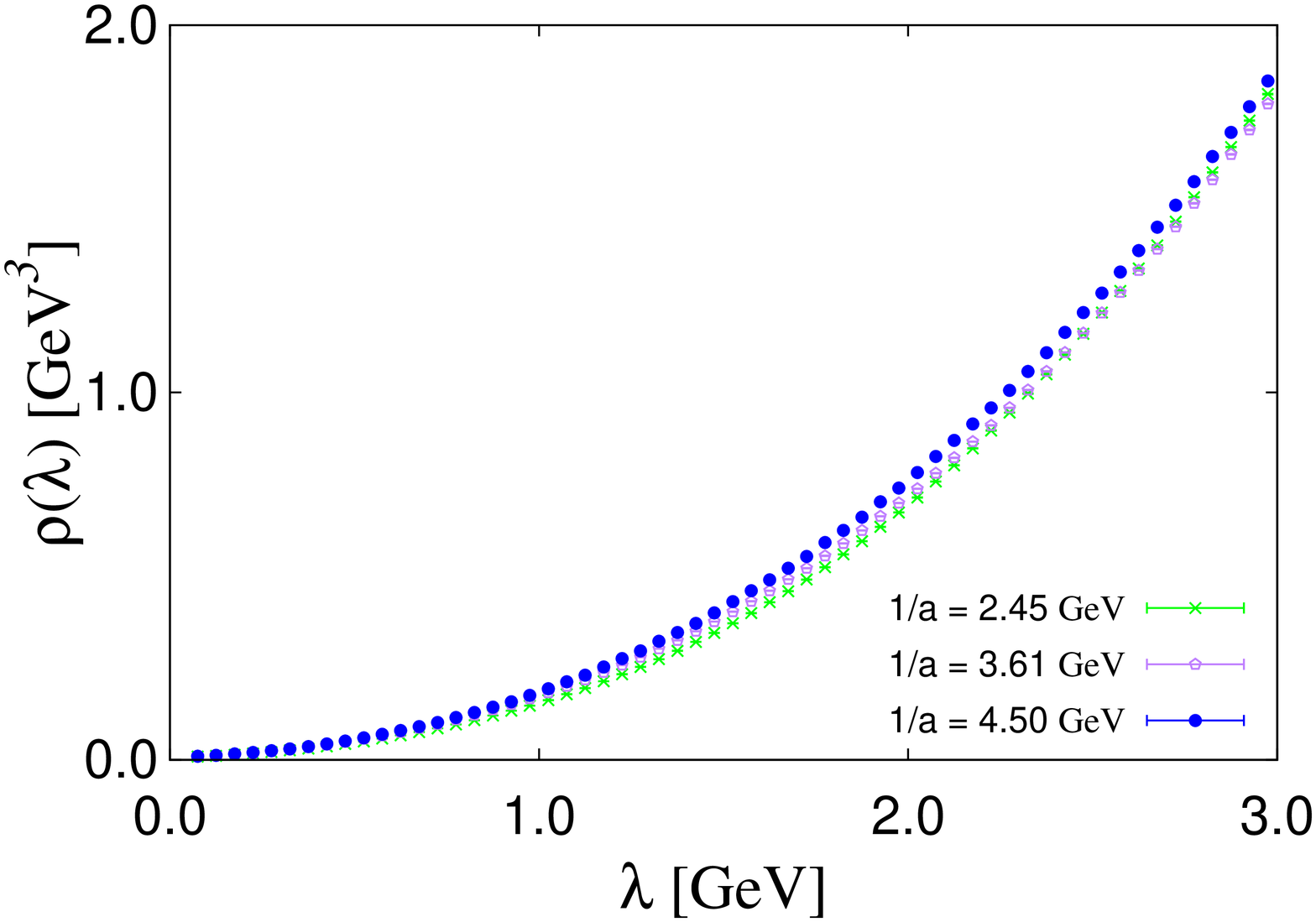}
  \caption{Dirac spectral density calculated at three lattice spacings, $1/a = 2.45$ GeV (cross), 3.61 GeV (circle), and 4.50 GeV (dot), with different Pauli-Villars mass $m_p = 1$ (left panel), and $m_p = 3$ (right panel).}
  \label{fig:gpv_dens}
\end{figure}

The domain-wall height $M_0$ corresponds to a large negative mass in $D_W$, and the kernel parameters $b$ and $c$ are chosen to realize good approximation of the sign function in finite fifth dimension. 
We use the standard choice $b - c = 1$, $M_0 = 1$, and $r = 1$, same as the Shamir type domain-wall fermion.
In this section, except for Figure \ref{fig:gpv_dens}, we take $L_s \rightarrow \infty$ for simplicity. It allows us to ignore the dependence on $b+c$ since the sign function becomes exact.

The free field case for the massless Dirac eigenvalue  may be written as $\lambda(m_p)\equiv\sqrt{D_{\mathrm{ov}} ^\dagger(0,m_p)D_{\mathrm{ov}}(0,m_p)}$ in the momentum space.
The massless Dirac eigenvalue of $D_\mathrm{OV}$ is bounded as $0\leq\lambda(m_p)\leq m_p$.

\begin{figure}[thb]
  \centering
  \sidecaption
  \includegraphics[width=4.5cm, angle=-90]{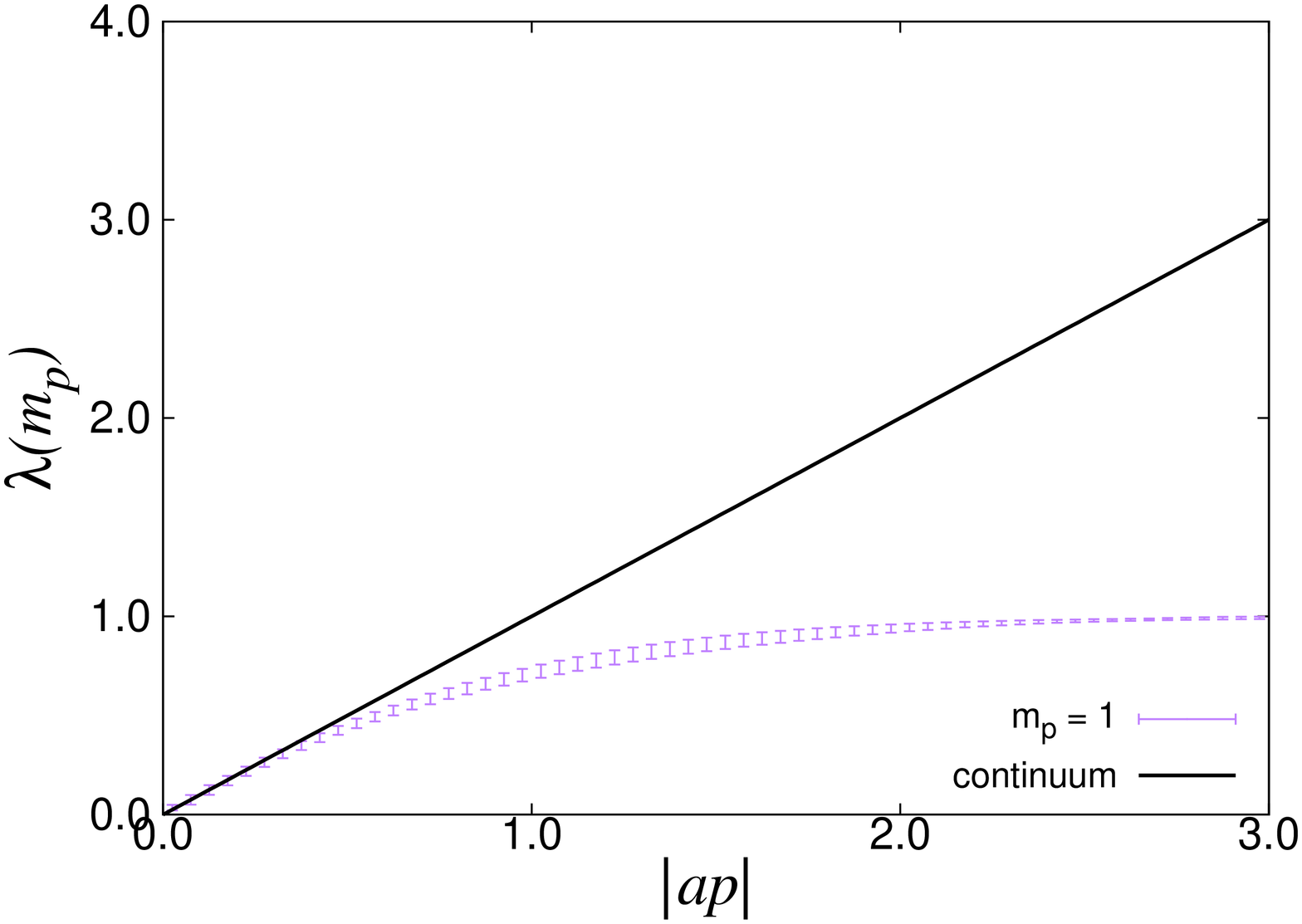}
  \includegraphics[width=4.5cm, angle=-90]{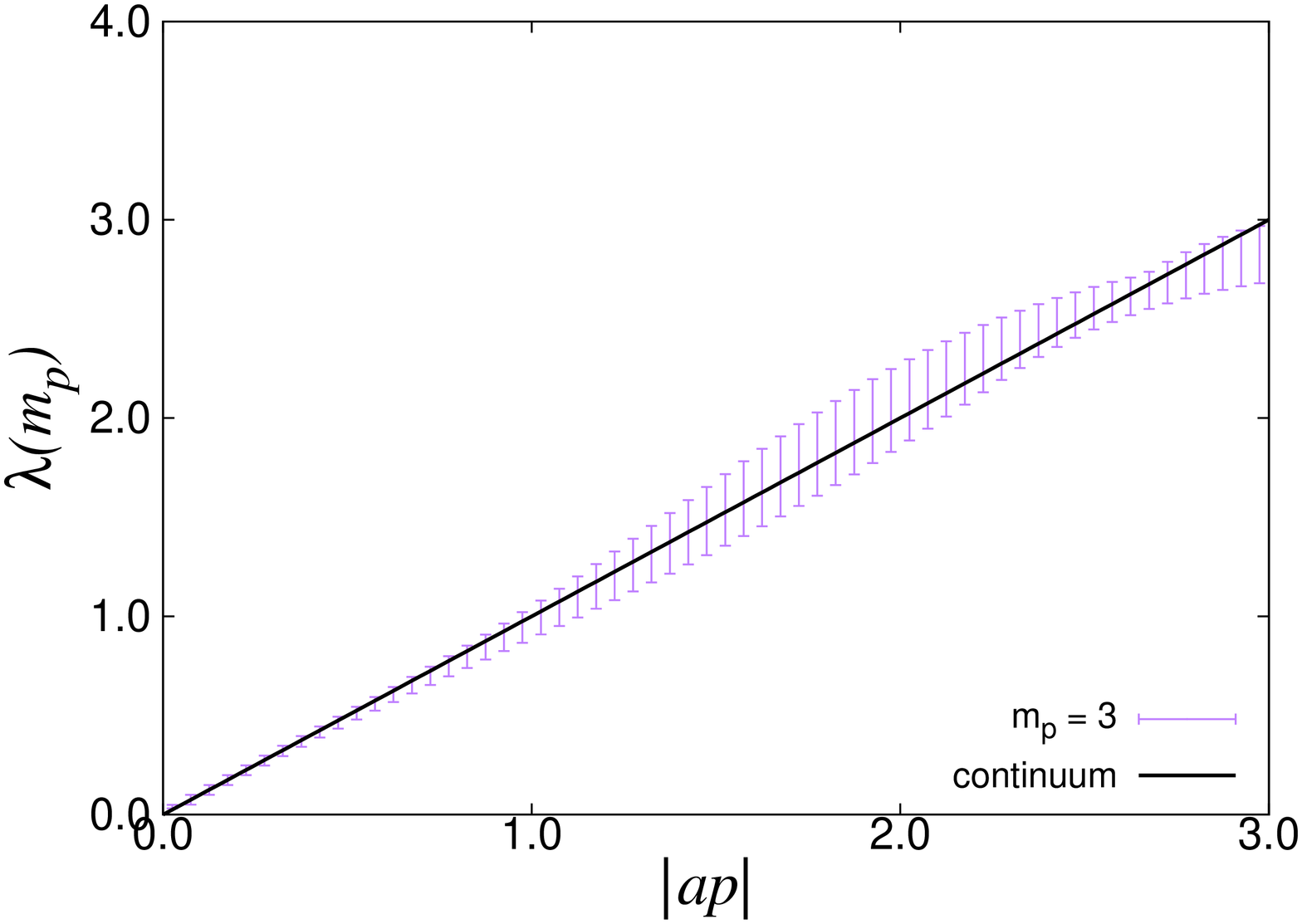}
  \caption{Eigenvalue of the Dirac operator on the trivial background, plotted as a function of the absolute value of momentum $|ap|$ with different Pauli-Villars mass $m_p = 1$ (left panel), and those of $m_p = 3$ (right panel).
  Bands show the minimum and maximum of the eigenvalue at each $|ap|$, since eigenvalue depends on the momentum direction:
  {\it e.g.} $ap_\mu = (2,0,0,0)$ and $ap_\mu = (1,1,1,1)$ have same $|ap|$, but have the different eigenvalues.}
  \label{fig:gpv_cutoff}
\end{figure}

Figure \ref{fig:gpv_cutoff} shows the eigenvalue $\lambda(m_p)$ as a function of the absolute value of the momentum $|ap|$.
The variation due to different momentum orientation for the same $|ap|$ is shown by the error bar.
We find a clear difference between the choices of $m_p=1$ (left panel), which corresponds to the standard domain-wall fermion, and another possible choice $m_p = 3$ (right panel).
It turned out that $m_p =3$ is much closer to the continuum relation $\lambda(m_p) = |ap|$. The difference may be written as 

\begin{equation}
  a^2\lambda^2(m_p)
  = \frac{a^2\lambda^2(1)}{
    1 + \left(\frac{1}{m_p^2} - 1\right) 
    \frac{a^2\lambda^2(1)}{(2-(b-c)M_0)^2M_0 ^2}},
\label{eq:gpv_rel}
\end{equation}
which is valid for arbitrary background gauge field,
since $D_{\mathrm{ov}}^\dagger D_{\mathrm{ov}}$'s with different
$m_p$ commute with each other.

We therefore choose $m_p =3$ for the study of the Dirac eigenvalue spectrum.
Since the ensembles are generated with $m_p=1$,
it corresponds to a partially quenched setup.
It is nevertheless harmless because the relation (\ref{eq:gpv_rel}) is one-to-one and it can be understood as a slightly modified observable.
In the continuum limit, the difference vanishes.
We note that the choice of $m_p$ has no effect on the pole structure of fermion propagators since $m_p$ changes only the denominator of (\ref{eq:Dov}) other than the overall scale.

Theoretically, $D_\mathrm{OV}(m_f,m_p)$ with $m_p \neq 1$ has the properties required for the overlap fermion such as the Ginsparg-Wilson relation, and the exponential locality.
The Ginsparg-Wilson relation is slightly modified depending on the Pauli-Villars mass,

\begin{equation}
  D_{\mathrm{ov}}^{-1}(0,m_p) \gamma_5 +
  \gamma_5 D_{\mathrm{ov}}^{-1}(0,m_p)
  =
  \frac{2a}{(2 - (b-c)M_0)M_0m_p}\gamma_5.
\end{equation}
The exponential locality is guaranteed along with the discussion in \cite{Hernandez:1998et}.
The key idea is that the sign function is not strictly local, but has an exponential locality, and $|\mathrm{sgn}(x)| = 1$. Then any polynomial of $\mathrm{sgn}(x)$ has an exponentially dumped upper bound.

In Figure \ref{fig:gpv_dens} we show the non-perturbative results on our generated lattice configurations at three lattice spacings.
The results with $m_p = 1$ (left panel) strongly depends on the lattice spacing, whereas the results with $m_p = 3$ (right panel) shows much milder dependence.

\section{Perturbative calculation of spectral density}\label{sec-2}

On the perturbative side, we construct the $O(\alpha_s ^4)$ coefficient in the $\overline{\mathrm{MS}}$ renormalization scheme for the exponent of the Dirac spectral density. In \cite{Kneur:2015dda}, an $O(\alpha_s ^3)$ calculation of the Dirac spectral density is given.
At the renormalization scale $\mu$ set to $\mu = \lambda(\mu = \lambda)$, it reads

\begin{equation}
\label{eq:ordert}
\rho(\mu = \lambda)
=
\frac{3\lambda^3}{4\pi^2}
\left(
1
+
1.06\alpha_s
-
2.14\alpha_s ^2
-
5.98\alpha_s ^3
+
O(\alpha_s ^4)
\right),
\end{equation}
for $n_f = 3$.
From now on, for simplicity, we sometimes suppress the renormalization scale dependence of $\lambda(\mu)$.

Now we focus on the exponent of $\rho(\lambda)$, ${\it i.e.}$
\begin{equation}
  F(\lambda) \equiv 
  \frac{\partial\ln\rho(\lambda)}{\partial\ln\lambda}.
  \end{equation}
Because an integral of the spectral density $\int_0^M d\lambda\,\rho(\lambda)$ with an upper limit $M$ is scale invariant
\cite{Giusti:2008vb},
the renormalization equation may be written as
\begin{equation}
  \label{eq:RG}
  0 = \left[
    \frac{\partial}{\partial\ln\mu}
    -\gamma_m(\alpha_s)
    \left(1+\lambda\frac{\partial}{\partial\lambda}\right)
    +\beta(\alpha_s)
    \frac{\partial}{\partial\alpha_s}
  \right]
  \rho(\lambda),
\end{equation}
 with the mass anomalous dimension $\gamma_m(\alpha_s)$ and the beta function $\beta(\alpha_s)$, defined as
\begin{equation}
  \beta(\alpha_s)  \equiv  
  \frac{\partial\alpha_s}{\partial\ln{\mu}},
  \label{eq:beta}
\end{equation}
\begin{equation}
  \gamma_m(\alpha_s)  \equiv  
  -\frac{\partial\ln m(\mu)}{\partial\ln\mu}.
  \label{eq:anodim}
\end{equation}
Here the beta function $\beta$ is known up to $O(\alpha_s ^6)$, and the mass anomalous dimension $\gamma_m$ is known to $O(\alpha_s ^5)$ \cite{Baikov:2016tgj,Baikov:2014qja}.
Using (\ref{eq:RG}) we can construct the $O(\alpha_s ^{i+1})$ exponent of the Dirac spectral density from the $O(\alpha_s ^i)$ Dirac spectral density since the beta funciton $\beta$ and mass anomalous dimension $\gamma_m$ start from $O(\alpha_s ^2)$ and from $O(\alpha_s)$, respectively, and both of them are known at least at $O(\alpha_s ^4)$.
We thus obtain
\begin{equation}
  \label{eq:F^MS}
  F^{\overline{\mathrm{MS}}}(\lambda) = 3 -
  F^{(1)}\frac{\alpha_s(\mu)}{\pi} -
  F^{(2)}\left(\frac{\alpha_s(\mu)}{\pi}\right)^2 -
  F^{(3)}\left(\frac{\alpha_s(\mu)}{\pi}\right)^3 -
  F^{(4)}\left(\frac{\alpha_s(\mu)}{\pi}\right)^4 
  + O(\alpha_s^5),
\end{equation}
with the coefficients $F^{(k)}$ for $n_f=3$ given as
\begin{eqnarray}
  \label{eq:F^(k)}
  F^{(1)} & = & 8,
  \\
  F^{(2)} & = & \frac{4}{3} \left(22 - 27 L_\lambda\right)
  \nonumber\\
          & = & 29.3333 - 36 L_\lambda,
  \\
  F^{(3)} & = & \frac{1}{36} \left( 6061 - 9216 L_\lambda + 
                5832 L_\lambda^2 - 1350 \pi^2 - 936 \zeta_3 \right)
  \nonumber\\
          & = & -233.003 - 256 L_\lambda + 162 L_\lambda^2,
  \\
  F^{(4)} & = & \frac{1}{5184} \left[
    \left( -3583861 + 1015200\pi^2 + 69984 c_3\pi^3 + 3888\pi^4
           -315168\zeta_3 - 432000\zeta_5 \right)
    \right.
  \nonumber\\
    & & \left.
    +\left(-10980576 + 2624400\pi^2 + 1819584\zeta_3\right) L_\lambda
    + 8771328 L_\lambda^2 - 3779136 L_\lambda^3\right]
  \nonumber\\
          & = & -1348.6655 + 3300.2425 L_\lambda + 1692 L_\lambda^2 -
                729 L_\lambda^3.
\end{eqnarray}
Here, $\zeta_5$ = 1.03692, 
$c_3\simeq 15993.5/(64\pi^3)-11292.4/(256\pi)$, 
and $L_\lambda \equiv \ln(\lambda/\mu)$.
At $\mu=\lambda(\mu = \lambda)$ it is numerically written as
\be{
 \label{eq:Fmu}
F(\lambda)_{\mu = \lambda}
=
3
-2.54648\alpha_s
-2.97209\alpha_s ^2
+7.51469\alpha_s ^3
+13.8454\alpha_s ^4
+O(\alpha_s ^5).
}

The leading order is $3$ as expected, because $\rho(\lambda)$ scales as $\lambda^{D-1}$ in $D$ dimensions.
We note that the expansion coincides with the relation $\rho(\lambda)\propto\lambda^{4/(1 + \gamma_m) - 1}$ up to the $O(\alpha_s)$ level, which is suggested for conformally invariant theories \cite{Patella:2012da,Cheng:2013eu}.
At the $O(\alpha_s)$ level, the mass anomalous dimension is $\gamma_m = 2\alpha_s/\pi$, then 
$\lambda^{4/(1 + \gamma_m) - 1} = \lambda^{3-8\alpha_s/\pi}$ at this order, while using (\ref{eq:F^MS},\ref{eq:F^(k)}), $\rho(\lambda)$ is proportional to $\lambda^{3 - F^{(1)}\alpha_s/\pi}$ with $F^{(1)} = 8$.
Beyond this order, such a simple relation between $F(\lambda)$ and $\gamma_m$ is lost because of non-zero $\beta$ function.

\section{Lattice result}\label{sec-3}
We calculate the Dirac spectral density stochastically on the lattice using (\ref{chiralcond}).
We set the bin size to $0.05$ GeV and the scale to $\mu = 2$ GeV on each lattice ensemble.
Our lattice emsembles are generated with $2+1$ flavor Mobius domain-wall fermion with lattice spacings $1/a = 2.45,3.61$, and $4.50$ GeV. Lattice size is chosen such that the physical volume is constant at about $2.6-2.8$ fm, {\it i.e.} $32^3\times64,48^3\times96$, and $64^3\times128$, respectively. Finite volume effects are negligible since our pion mass is $230-500$ MeV in this calculation.
The Dirac spectral density has only tiny sea quark mass dependence, which can be safely ignored.
Our calculation is done with the Pauli-Villars mass $m_p = 1$, and the results are transformed to $m_p = 3$ by (\ref{eq:gpv_rel}).
To match our lattice calculation with the continuum $\overline{\mathrm{MS}}$ scheme, we use the renormalization constant determined through the short-distance vacuum polarization function analysis \cite{Tomii:2017cbt}.

\begin{figure}[thb]
  \centering
  \sidecaption
  \includegraphics[width=4.5cm, angle=-90]{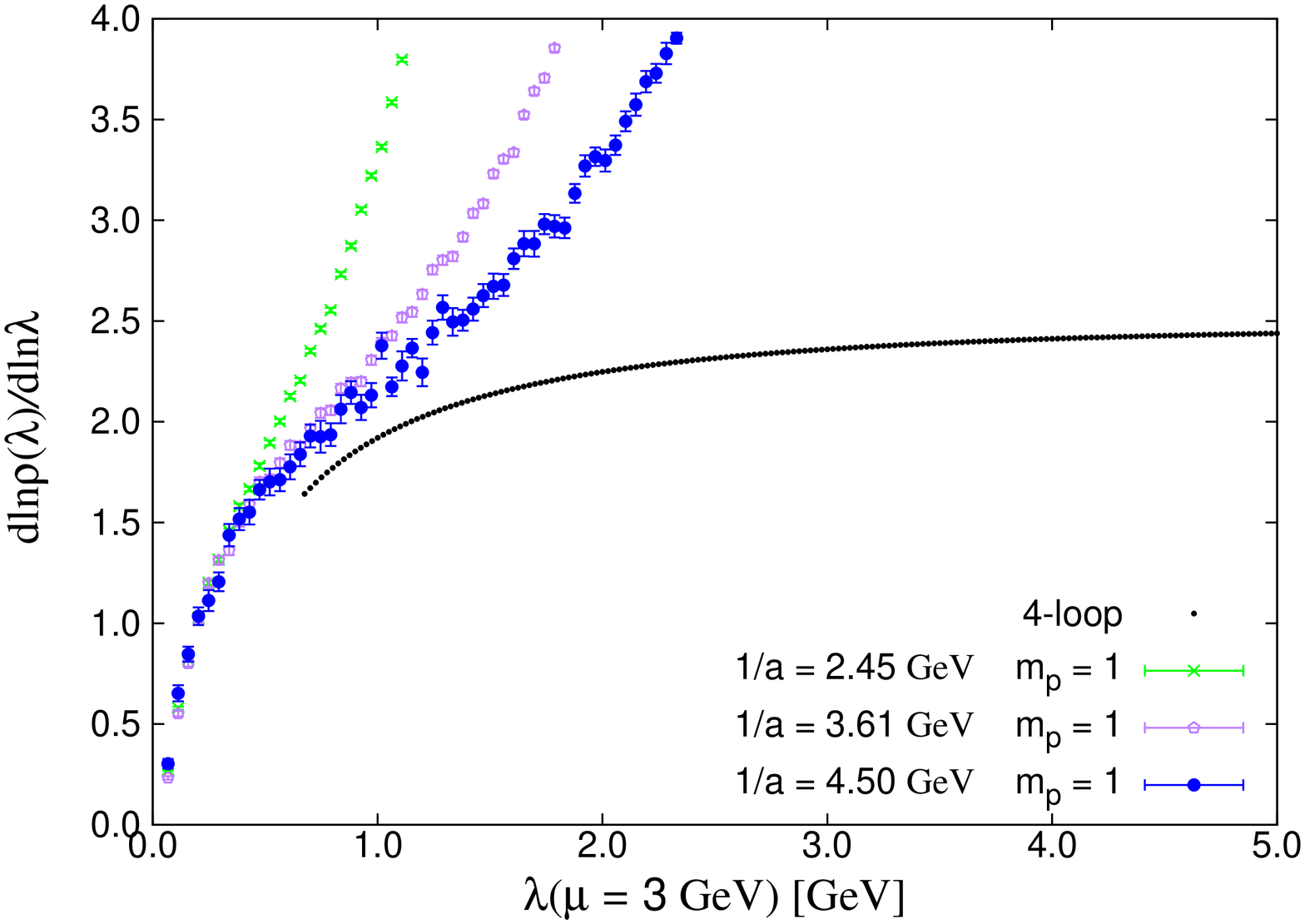}
  \includegraphics[width=4.5cm, angle=-90]{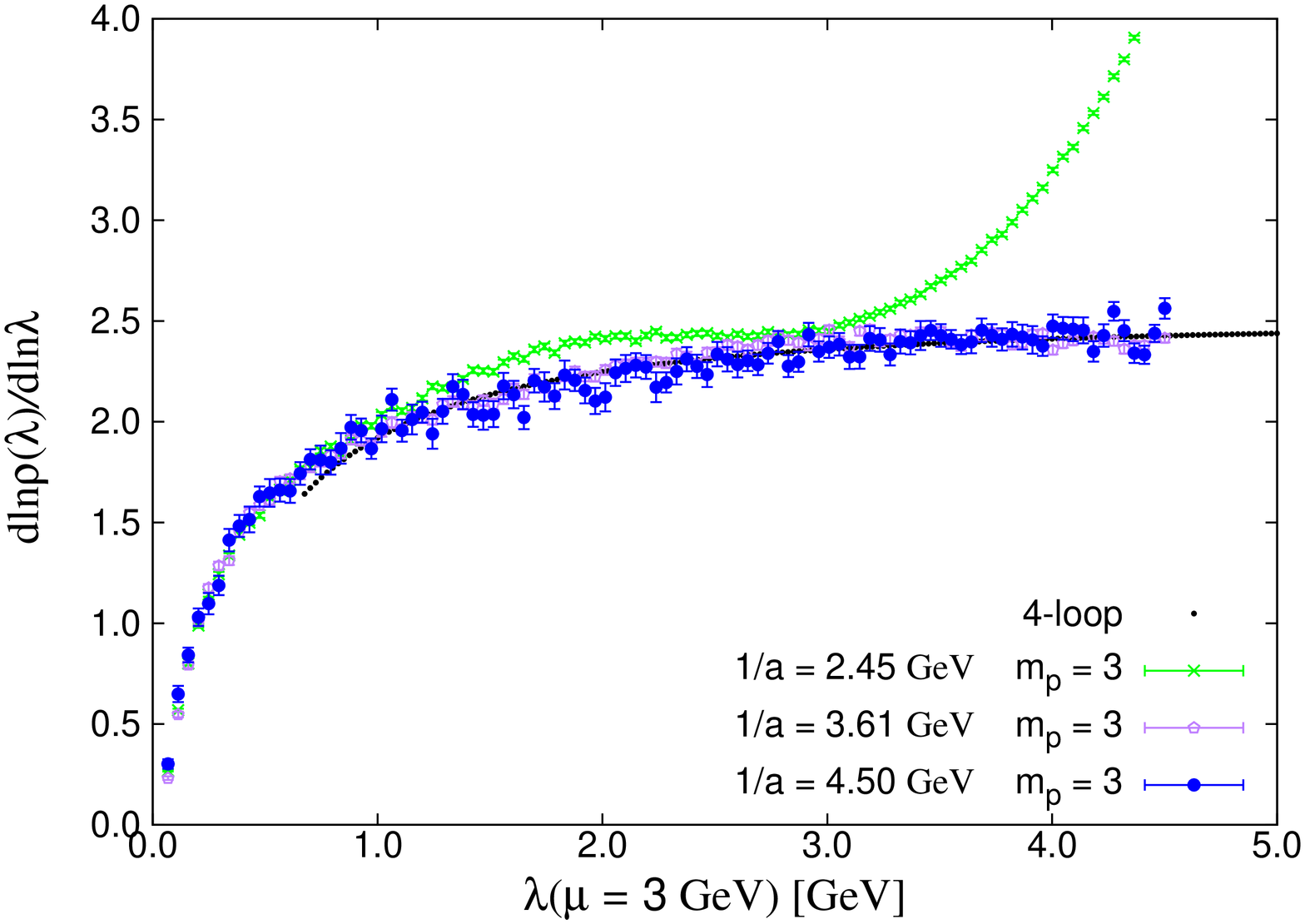}
  \caption{Exponent of the Dirac spectral density calculated at, $1/a = 2.45$ GeV (cross), 3.61 GeV (circle), and 4.50 GeV (dot). The perturbative calculation is also shown (dotted line) with different Pauli-Villars masses $m_p = 1$ (left panel), and $m_p = 3$ (right panel).}
  \label{fig:gpv_expo}
\end{figure}

\begin{figure}[thb]
  \centering
  \sidecaption
  \includegraphics[width=8cm, angle=-90]{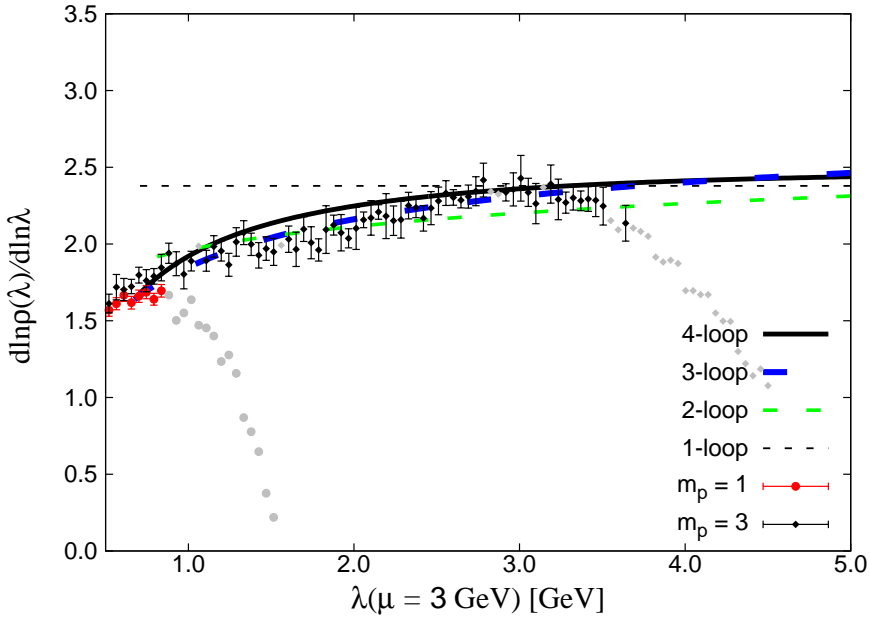}
  \caption{Exponent of the Dirac spectral density after the continuum extrapolation $m_p = 1$ (dot) and $m_p = 3$ (square).
  Perturbative calculations at one-loop order (thin dashed line) as well as at higher orders (thicker line) are also plotted.
  }
  \label{fig:gpv_extract}
\end{figure}

Figure \ref{fig:gpv_expo} shows the exponent of the Dirac spectral density calculated at each lattice spacing compared with the four-loop perturbative calculation.
As we already discussed for the spectral density, the Pauli-Villars mass $m_p = 3$ reduces the discretization effect, and enables us to extrapolate to the continuum limit with much smaller uncertainty.
After the extrapolation, the exponent of the Dirac spectral density is shown in Figure \ref{fig:gpv_extract}.
The error bar represents the sum of the statistical error, the uncertainty from lattice spacings, and from renormalization constant in quadrature.
Gray dots without error bars have $\chi^2/$d.o.f $>2$, {\it i.e.} the continuum limit in this region are not reliable because of the large lattice artifact.
In particular, the results with $m_p = 1$ are difficult to extrapolate for the eigenvalues $\lambda \geq 1$ GeV, because of the large discritization effects.
Those with $m_p = 3$ are reliable up to $\lambda \simeq 3$ GeV.

We also pay attention to the reliability of the perturbative calculation depending on the scale of the eigenvalue.
We estimate the error from truncation of the perturbative expansion by using a difference between 3-loop and 4-loop results.
It should be minimal for $\lambda \simeq \mu$ since the perturbative calculation is written with $L_\lambda = \ln{\lambda/\mu}$.

Figure \ref{fig:gpv_extract} clearly shows that the difference is small around $\lambda =3$ GeV, which is our choice for the renormalization scale $\mu = 3$ GeV, while the larger error is observed around $\lambda \simeq 2$ GeV.

In order to see the argeement quantitatively, we choose the bin at $\lambda = 2.92$ GeV, as an example, close to $\mu = 3$ GeV.
The exponent of the Dirac specral density is obtained as $\mathrm{dln}\rho(\lambda)/\mathrm{dln}\lambda = 2.34(4)(2)(3)$, where the statistical error, the uncertainty originated from the lattice spacings, and the renormalization constant are shown as the estimated errors.
It is consistent with the $O(\alpha_s ^4)$ perturbative calculation, $\mathrm{dln}\rho(\lambda)/\mathrm{dln}\lambda = 2.35(1)$.

Eigenvalue density in the high energy region can be calculated perturbatively, while it is also accessible by lattice calculation.
We test the perturbation theory by the lattice calculation after $m_p = 3$ an extrapolation to the continuum limit.
The precision of $\sim 5\%$ is obtained for the exponent of the Dirac spectral density.

The result may also be used for the determination of the strong coupling constant $\alpha_s$, an analysis of which is underway.

\section*{Acknowledgements}

The lattice QCD simulation has been performed on Blue Gene/Q supercomputer at the High Energy Accelerator Research Organization (KEK) under the Large Scale Simulation Program (Nos. 15/16-09, 16/17-14). This work is supported in part by the Grant-in-Aid of the Japanese Ministry of Education (No. 26247043, 26400259).



\end{document}